\documentclass[10pt,twocolumn, nofootinbib]{revtex4-2}

\usepackage{assumptionsofphysics}
\usepackage{graphicx}
\usepackage{hyperref}
\hypersetup{
	colorlinks=true,
	citecolor=blue,
	urlcolor=blue,
	linkcolor=blue
}
\begin{document}

\title{Geometric and physical interpretation of the action principle}
\author{Gabriele Carcassi, Christine A. Aidala}
\affiliation{Physics Department, University of Michigan, Ann Arbor, MI 48109}

\date{\today}

\begin{abstract}
We give a geometric interpretation for the principle of stationary action in classical Lagrangian particle mechanics. In a nutshell, the difference of the action along a path and its variation effectively ``counts'' the possible evolutions that ``go through'' the area enclosed. If the path corresponds to a possible evolution, all neighbouring evolutions will be parallel, making them tangent to the area enclosed by the path and its variation, thus yielding a stationary action. This treatment gives a full physical account of the geometry of both Hamiltonian and Lagrangian mechanics which is founded on three assumptions: determinism and reversible evolution, independence of the degrees of freedom and equivalence between kinematics and dynamics. The logical equivalence between the three assumptions and the principle of stationary action leads to a much cleaner conceptual understanding.
\end{abstract}

\maketitle

\section{Introduction}

While the principle of stationary action is regarded by many as one of the most important tools in physics, its physical meaning is not completely clear.\cite{hamilton1834general, heaviside1903, feynman1942principle, brenier1989least, rojo2018principle} First of all, the typical characterization of the Lagrangian as the difference between kinetic and potential energy fails even for simple systems, like a charged particle under a magnetic field. Moreover, the Lagrangian for a system is not uniquely defined, making the actual value of the action for a path not directly physically significant.  We are left to wonder: what exactly is the action and why is it stationary for actual trajectories?

As part of our larger project Assumptions of Physics, we developed an approach, called Reverse Physics \cite{aop-phys-ReversePhysics}, which examines current theories to find a set of starting physical assumptions that are sufficient to rederive them. We have found that Lagrangian mechanics is equivalent to three assumptions: determinism/reversibility, independence of degrees of freedom and kinematics/dynamics equivalence \cite{aop-phys-blueprint}. This physically motivated understanding of the classical theory can be used to characterize both the physics and the geometry underlying the principle of stationary action. What we find is that this arises as a general mathematical feature of divergence-free fields (and closed two-forms), which are the appropriate tools to describe a flow that conserves the number of states. The assumption of equivalence between kinematics and dynamics (i.e. we can reconstruct the dynamical state simply by looking at the trajectory) is what allows us to express the principle in the usual form. The argument can proceed in the reverse direction: assuming the principle of stationary action recovers a dynamical system that exhibits those three physical assumptions.

The mathematics needed to run the argument is well established\cite{souriau1970structure, abraham1978foundations,arnold1989mathematical, marsden1999introduction}. The two aspects that have been sorely lacking, and that we provide, are (1) a clear geometric interpretation of the action principle and (2) a tight connection between the math and the physics it represents. [This has been the hardest problem to solve, as the modern mathematical concepts and notation have departed from physical intuition. We will therefore use, instead of the standard notation of differential geometry, an extended version of the one physicists use in general relativity.] To make the result accessible to the widest audience, we first cover the case of a single degree of freedom using standard vector calculus. We then proceed to the case of multiple independent degrees of freedom using tools from differential geometry. The main article will present all the key points needed to follow the argument and its physical and geometric meaning, leaving the mathematical details and calculations to the Supplementary Information.

\section{One degree of freedom}

\begin{figure}
	\includegraphics[width = 0.45\textwidth]{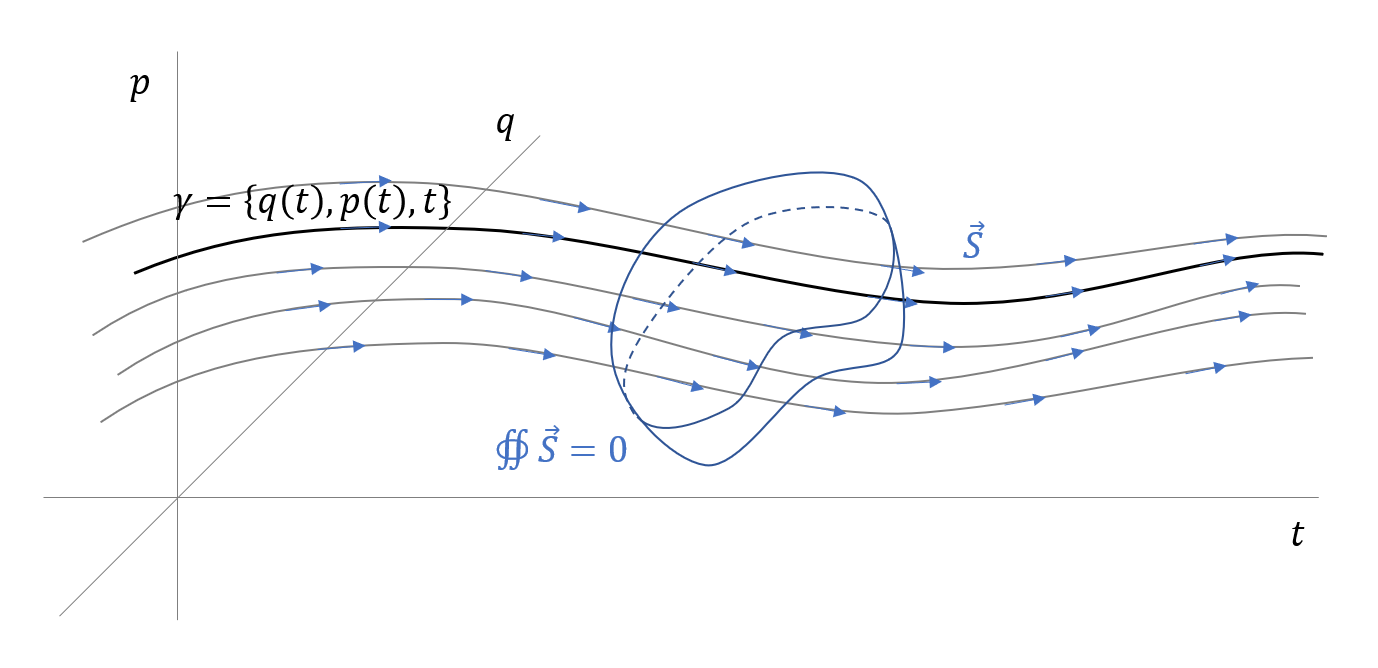}
	\caption{\footnotesize{Evolutions in the extended phase space and the divergence-free displacement field.}}\label{extended_phase_space}
\end{figure}

As we want to characterize the evolution of states over time, the appropriate setting is phase space extended with the time variable\cite{lanczos1949variational,synge1960encyclopedia}. That is, the space charted by position $q$, momentum $p$ and time $t$ as can be seen in Fig. \ref{extended_phase_space}. In the same way that we write $x^i = [ x, y, z ]$ for the three dimensions of space, we write
\begin{equation}\label{sdof_variables}
	\xi^a = [ q, p, t]
\end{equation}
for the three dimensions of the extended phase space.

Under the assumption that
\begin{align}\label{assum_detrev}
	\tag*{(DR)}
	\parbox{2.8in}{\emph{the system undergoes deterministic and reversible evolution}}
\end{align}
we can define a displacement vector field
\begin{equation}\label{sdof_displacement}
	\begin{aligned}
		\vec{S} &= \left[ \frac{dq}{dt},\frac{dp}{dt},\frac{dt}{dt} \right] \\
		&= S^a e_a = \frac{d\xi^a}{dt} e_a .
	\end{aligned}
\end{equation}
that describes how states move in time.\footnote{Where possible, we will be writing the same expression in both vector calculus and component notations.} In dynamical system literature, this is referred to as the vector field of the dynamical system. The time component of the displacement vector field is constrained, as we have
\begin{equation}\label{sdof_time_constraint}
	S^t=\frac{dt}{dt}=1.
\end{equation}

If assumption \ref{assum_detrev} is valid, we expect the flow of states through a closed surface to be zero: as many states flow in as flow out of the region. Alternatively, if we assign a probability, or probability density, to each trajectory, the assumption requires that probability not to change, so integrating the probability over a closed surface must yield zero. However we see it, assumption \ref{assum_detrev} means the field is divergence-free.\footnote{Given that this is a three-dimensional space, we can use the standard tools of vector calculus.} That is, 
\begin{equation}\label{sdof_div_free}
	\nabla \cdot \vec{S} = \partial_a S^a = 0.
\end{equation}

Since the displacement field is divergence-free, it admits a vector potential. We have
\begin{equation}\label{sdof_displacement_potential}
	\begin{aligned}
		\vec{\theta} &= [\theta_q, \theta_p, \theta_t] = \theta_a e^a \\
		\vec{S} &= - \nabla \times \vec{\theta} = - \epsilon^{abc} \partial_b \theta_c \, e_a. \\
	\end{aligned}
\end{equation}
The minus sign is introduced to match conventions. Mathematically, this is analogous to what is done for a magnetic field or for an incompressible fluid.

Because the displacement field must satisfy \ref{sdof_time_constraint}, without loss of generality we can set
\begin{equation}\label{sdof_potential_expression}
	\begin{aligned}
		\vec{\theta} &= [p, 0, -H(q,p,t)] \\
		&= p e^q - H(q,p,t) e^t,
	\end{aligned}
\end{equation}
where $H$ is a suitable function of $q$, $p$ and $t$. The potential $\vec{\theta}$ is closely related to the canonical one-form of symplectic geometry and the contact form of contact geometry. By applying definition \ref{sdof_displacement} and expanding \ref{sdof_displacement_potential} with \ref{sdof_potential_expression}, we have
\begin{equation}\label{sdof_Ham_eq}
	\left[ \frac{dq}{dt},\frac{dp}{dt},\frac{dt}{dt} \right] = - \nabla \times \vec{\theta} = \left[ \frac{\partial H}{\partial p},-\frac{\partial H}{\partial q}, 1 \right],
\end{equation}
which yields Hamilton's equations. Note that the argument works in reverse: any Hamiltonian system with one degree of freedom yields a divergence-free displacement field, and therefore satisfies \ref{assum_detrev}.

\begin{figure}
	a \includegraphics[width = 0.45\textwidth]{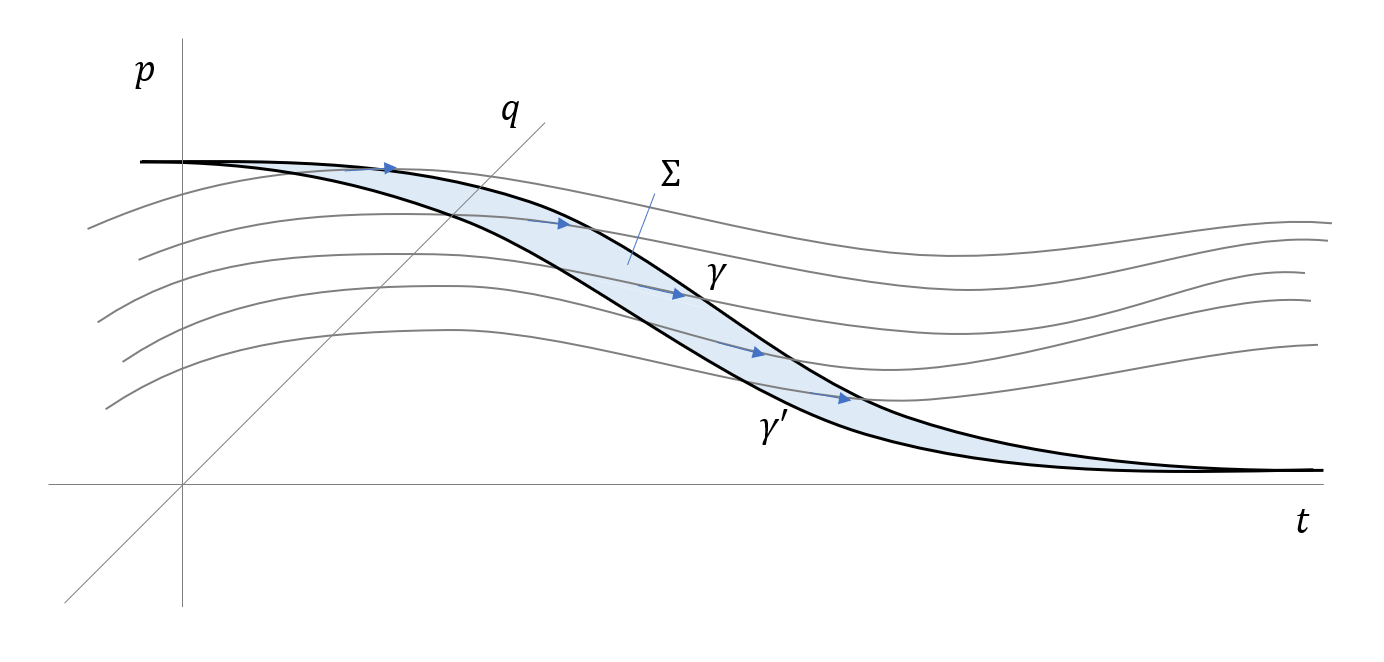} \\
	b \includegraphics[width = 0.45\textwidth]{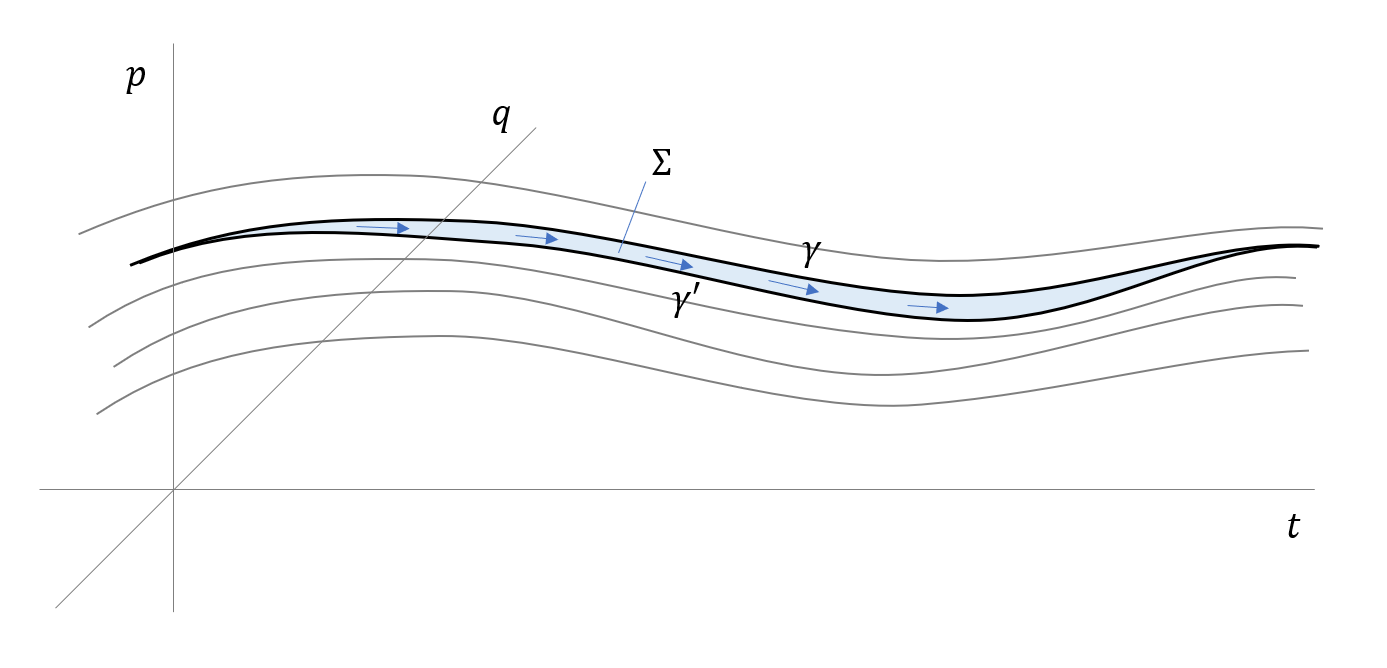}
	\caption{\footnotesize{The variation of the action is the flow of the displacement field $\vec{S}$ through the surface $\Sigma$ that sits between the path $\gamma$ and its variation $\gamma'$. In (b) we see that the flow is zero if the path is an actual evolution of the system, since the displacement field will be parallel to the path $\gamma$ and therefore tangent to the surface $\Sigma$.}}\label{fig_action}
\end{figure}

We now turn to constructing the principle of stationary action. As illustrated in Fig. \ref{fig_action}a, take a path $\gamma$ with endpoints $A$ and $B$, not necessarily a solution of the equations of motion. Then take a variation $\gamma'$ of that path and identify a surface $\Sigma$ between them. We can ask: what is the flow of the displacement field $\vec{S}$ through $\Sigma$? Because $\vec{S}$ is divergence-free, the flow through $\Sigma$ will depend only on the contour, therefore the question is well posed. Using Stokes' theorem, we find
\begin{equation}\label{sdof_action}
	\begin{aligned}
		- \iint_{\Sigma} \vec{S} \cdot d\vec{\Sigma} &= \iint_{\Sigma} \left( \nabla \times \vec{\theta} \right) \cdot d\vec{\Sigma} \\
		&=  \oint_{\partial \Sigma = \gamma \cup \gamma'} \vec{\theta}  \cdot d\vec{\gamma}  \\
		&= \int_{\gamma} \vec{\theta} \cdot d\vec{\gamma} - \int_{\gamma'} \vec{\theta} \cdot d\vec{\gamma}' \\
		&= \delta \int_{\gamma} \vec{\theta} \cdot d\vec{\gamma}.
	\end{aligned}
\end{equation}

Now suppose $\gamma$ is a solution of the equation of motion, as in Fig. \ref{fig_action}b. Then $\gamma$ is a field line and the flow is tangent to $\Sigma$ no matter what $\gamma'$ we picked. The converse is true: if we look for those paths for which the flow through $\Sigma$ is zero no matter what $\gamma'$, $\gamma$ must be everywhere tangent to $\vec{S}$ so we find a solution to the equation of motion. The solutions, then, are those paths and only those paths for which
\begin{equation}\label{sdof_stationary_action}
	0 =\delta \int_{\gamma} \vec{\theta} \cdot d\vec{\gamma} = - \iint_{\Sigma} \vec{S} \cdot d\vec{\Sigma} 
\end{equation}
We call this the principle of stationary action in Hamiltonian form.

The last step is to express the principle exclusively in terms of kinematic variables: position, time and velocity. This can be done if we assume that 
\begin{align}\label{assum_kineq}
	\tag*{(KE)}
	\parbox{2.8in}{\emph{the kinematics of the system is enough to reconstruct its dynamics}.}
\end{align}
This means that by looking at just the trajectory in space $q(t)$, we are able to reconstruct the state at each moment in time. Therefore we must be able to write $p=p(q,\dot{q})$, and therefore we can also write
\begin{equation}\label{sdof_Lagrangian}
	\begin{aligned}
		\delta \int_{\gamma} \vec{\theta} \cdot d\vec{\gamma} 
		&= \delta \int^{t_2}_{t_1} \vec{\theta} \cdot \frac{d\vec{\gamma}}{dt} dt \\  
		&= \delta \int^{t_2}_{t_1} \left(p \frac{dq}{dt} - H \right) dt \\
		&= \delta \int^{t_2}_{t_1}L(q, \dot{q}, t) dt = 0.
	\end{aligned}
\end{equation}

We find that a system for which \ref{assum_detrev} and \ref{assum_kineq} are valid can be characterized in terms of the principle of stationary action with a suitable Lagrangian. The converse is also true: if the principle of stationary action allows for a unique solution, then the conjugate momentum and the Hamiltonian are well defined and the system satisfies both \ref{assum_detrev} and \ref{assum_kineq}.

We have thus demystified the principle of stationary action, and turned it into a geometric property: requiring the principle of stationary action is equivalent to requiring that the solutions are the field lines of a divergence-free field in phase space. We also have a clear physical meaning: the principle of stationary action is equivalent to assuming determinism/reversibility \ref{assum_detrev} and kinematic equivalence \ref{assum_kineq}. However, we do feel that the principle expresses these requirements in a very roundabout way.

\section{Multiple degrees of freedom}

To generalize to multiple degrees of freedom we have to abandon the tools provided by vector calculus and embrace the ones provided by differential geometry. For $N$ independent degrees of freedom, we will have the $2N+1$ manifold charted by the variables
\begin{equation}\label{mdof_variables}
	\xi^a = [ q^i, p_i, t].
\end{equation}
The displacement vector field will be
\begin{equation}\label{mdof_displacement}
	\begin{aligned}
		\vec{S} &= S^a e_a = \frac{d\xi^a}{dt} e_a =\frac{dq^i}{dt} e_{q^i} + \frac{dp_i}{dt} e_{p_i} + \frac{dt}{dt} e_t
	\end{aligned}
\end{equation}
which also satisfies \ref{sdof_time_constraint}.

The flow of $\vec{S}$ will still be divergence-free, but this property only tells us that the total number of states is conserved. We also need to assume that
\begin{align}\label{assum_indep}
	\tag*{(IND)}
	\parbox{2.8in}{\emph{the degrees of freedom are independent.}}
\end{align}
This means that, at each moment in time, the total number of states (i.e. the volume) of a parallelepiped will be the product of states identified by each degree of freedom (i.e. the areas). To be able to quantify the number of states identified by each degree of freedom, we introduce the rank-2 tensor (more precisely a two-form)
\begin{equation}\label{mdof_form}
	\omega = \omega_{ab} \, e^a \otimes e^b,
\end{equation}
which we call the state counting form.\footnote{Note that this approach is consistent with statistical mechanics\cite{peliti2011statistical}.}

The counting form must be of rank two as independent degrees of freedom are bi-dimensional (i.e. quantity plus conjugate). Given two vectors $\vec{v}$ and $\vec{w}$, these identify a parallelogram in phase space and 
\begin{equation}\label{mdof_form_applied}
	\omega(\vec{v}, \vec{w}) = \omega_{ab} v^a w^b
\end{equation}
quantifies the number of states over its surface.

The form $\omega$ will need to be anti-symmetric
\begin{equation}\label{mdof_form_antisymm}
	\omega(\vec{v}, \vec{w}) = - \omega(\vec{w}, \vec{v})
\end{equation}
as the parallelogram identified by $\vec{v}$ and $\vec{w}$ in that order will be the same as the one identified by $\vec{w}$ and $\vec{v}$ with opposite orientation. The form $\omega$ will also be closed, meaning 
\begin{equation}\label{mdof_closed_form}
	\oiint_\Sigma \omega(d\Sigma) = 0
\end{equation}
over all closed contractible surfaces. [In our notation, $d\Sigma$ represents the infinitesimal surface element of integration, which is the argument of the form.] This stems from \ref{assum_indep}. Suppose you take a surface and translate it along another independent variable. The count of states cannot depend on the independent variable. Therefore if we construct any parallelepiped at equal time, two opposite sides will contain the same number of states with opposite orientation, and therefore the integral of $\omega$ over the whole surface will be zero. Those familiar with differential geometry will recognize $\omega$ as the symplectic form of symplectic manifolds and of the symplectic bundles of contact manifolds.

Since $\omega$ is closed, in every contractible region it can be expressed as the exterior derivative of a covector $\theta$:
\begin{equation}\label{mdof_form_potential}
	\begin{aligned}
		\theta &= \theta_a e^a = \theta_{q^i} e^{q^i} + \theta_{p_i} e^{p_i} + \theta_t e^t \\
		\omega &= - \partial \wedge \theta = - \left( \partial_a \theta_b - \partial_b \theta_a \right) e^a \otimes e^b
	\end{aligned}
\end{equation}
The minus sign is required to match conventions.\footnote{We write the exterior derivative as $\partial \wedge \omega$ instead of $d \omega$ so that $d$ is only used for infinitesimal line or surface elements.} 

The above is the generalization of \ref{sdof_displacement_potential} and it is important to note the differences. The exterior derivative takes the place of the curl and it is $\omega$, not $\vec{S}$, that can be expressed in terms of the potential $\theta$. Yet, there is an important geometric relationship we still need to take into account: the direction of the displacement does not contribute new states. Physically, this is a consequence of \ref{assum_detrev}: states cannot be created or disappear over time. We have
\begin{align}\label{mdof_displacement_kills}
	\omega(\vec{S}, \cdot) = 0.
\end{align}
Mathematically we say that the displacement vector kills the form; it makes it zero no matter what vector we put on the other side.

Similarly to the single d.o.f., without loss of generality we can set
\begin{equation}\label{mdof_potential_expression}
	\begin{aligned}
		\theta &= [p_i, 0, -H] \\
		&= p_i e^{q^i} - H e^t.
	\end{aligned}
\end{equation}
and calculate the components of the counting form
\begin{equation}\label{mdof_form_components}
	\omega_{ab} = (-\partial\wedge\theta)_{ab} = \begin{bmatrix}
		0 & \delta^j_i & \partial_{q^i} H \\
		-\delta^i_j & 0 & \partial_{p_i} H \\
		-\partial_{q^j} H & -\partial_{p_j} H & 0
	\end{bmatrix}.
\end{equation}

By combining \ref{mdof_displacement}, \ref{mdof_displacement_kills} and \ref{mdof_form_components} one finds
\begin{equation}\label{mdof_Ham_eq}
	\begin{aligned}
		S^{q^j} &= \partial_{p_j} H \\
		S^{p_j} &= - \partial_{q^j} H
	\end{aligned}
\end{equation}
which recovers Hamilton's equations.

The setup to recover the principle of stationary action is the same, except that we need to substitute vector calculus with differential geometry. First we find that the flow through $\Sigma$ is the variation of the action using Stokes' theorem
\begin{equation}\label{mdof_action}
	- \iint_{\Sigma} \omega(d\Sigma) = \delta \int_{\gamma} \theta(d\gamma) 
\end{equation}
If $\gamma$ is a solution of the equations of motion, then $d\gamma = \vec{S} dt$. The infinitesimal surface $d\Sigma$ will be a parallelogram formed by $d\gamma$ and the direction of the variation $d\lambda$. By \ref{mdof_displacement_kills}, $\vec{S}$ is the only vector field that kills the form $\omega$ therefore the solutions are such that
\begin{equation}\label{mdof_stationary_action}
	0 = \delta \int_{\gamma} \theta(d\gamma) = - \iint_{\Sigma} \omega(\vec{S}, d\lambda) dt. 
\end{equation}

Adding assumption \ref{assum_kineq}, we can express the variational principle in the usual form
\begin{equation}\label{mdof_Lagrangian}
	\begin{aligned}
		\delta \int_{\gamma} \theta(d\gamma) &= \delta \int^{t_2}_{t_1}\left( p_i \frac{dq_i}{dt} - H \right) dt \\
		&= \delta \int^{t_2}_{t_1}L(q^i, \dot{q}^i, t) dt = 0.
	\end{aligned}
\end{equation}

\section{Discussion and conclusions}

We have found that the principle of stationary action is equivalent to assumptions \ref{assum_detrev}, \ref{assum_indep} and \ref{assum_kineq} and that the variation of the action can be understood as the flow of evolutions through the surface delimited by the path and its variation. This allows us to fully understand both the physical and geometric significance of the principle in a much more precise way.

Note that the true physical content is in the displacement field $\vec{S}$ and the counting form $\omega$ whose properties descend from the assumptions. The action, the Lagrangian and the potential $\theta$ are not uniquely defined and are subject to a convenient, though still arbitrary, choice of gauge. The Lagrangian and the action, then, do not directly encode physical properties of the system. Saying that ``nature chooses the path that minimizes the action'', therefore, does not provide clear insight. On the other hand, the three assumptions give us a clear physical picture.

Another interesting element is that a version of the principle of stationary actions holds even if \ref{assum_kineq} does not apply. In this case the ``Lagrangian'' $L(q,\dot{q},p,t)$ depends on conjugate momentum as well. That is, the principle holds on all Hamiltonian systems, whether or not they admit a Lagrangian. Conversely, the cases where the Lagrangian does not admit a Hamiltonian are exactly the cases where the principle of stationary action fails to yield a unique solution, thus does not hold. These are the cases where the Lagrangian is not hyperregular. Therefore the principle of stationary action holds for all Lagrangian systems that admit a Hamiltonian, but also for all Hamiltonian systems whether of not they admit a Lagrangian. In this sense, the action principle is better understood as a feature of Hamiltonian mechanics, instead of Lagrangian mechanics\cite{souriau1970structure, arnold1989mathematical}.

As a final comment, note that privileging the Hamiltonian picture is more in line with quantum mechanics; the idea of $\omega$ as a tool to count states is in line with statistical mechanics; expression \ref{mdof_potential_expression} is remarkably similar to that of the relativistic four-momentum. This is one of the key insights we get from our Reverse Physics approach: there is a unity among the different physical theories that begs to be brought to light. We are convinced that a version of this geometric understanding must exist in the quantum world.

\vspace{-1mm}
\section*{Acknowledgments}
This paper forms part of the ongoing \textit{Assumptions of Physics} project \cite{aop-book}, which aims to identify a handful of physical principles from which the basic laws can be rigorously derived. The authors would like to thank Alejandro Uribe and Daniel Burns for helpful discussions. 

\bibliography{bibliography}

\begin{thebibliography}{15}%
\makeatletter
\providecommand \@ifxundefined [1]{%
 \@ifx{#1\undefined}
}%
\providecommand \@ifnum [1]{%
 \ifnum #1\expandafter \@firstoftwo
 \else \expandafter \@secondoftwo
 \fi
}%
\providecommand \@ifx [1]{%
 \ifx #1\expandafter \@firstoftwo
 \else \expandafter \@secondoftwo
 \fi
}%
\providecommand \natexlab [1]{#1}%
\providecommand \enquote  [1]{``#1''}%
\providecommand \bibnamefont  [1]{#1}%
\providecommand \bibfnamefont [1]{#1}%
\providecommand \citenamefont [1]{#1}%
\providecommand \href@noop [0]{\@secondoftwo}%
\providecommand \href [0]{\begingroup \@sanitize@url \@href}%
\providecommand \@href[1]{\@@startlink{#1}\@@href}%
\providecommand \@@href[1]{\endgroup#1\@@endlink}%
\providecommand \@sanitize@url [0]{\catcode `\\12\catcode `\$12\catcode
  `\&12\catcode `\#12\catcode `\^12\catcode `\_12\catcode `\%12\relax}%
\providecommand \@@startlink[1]{}%
\providecommand \@@endlink[0]{}%
\providecommand \url  [0]{\begingroup\@sanitize@url \@url }%
\providecommand \@url [1]{\endgroup\@href {#1}{\urlprefix }}%
\providecommand \urlprefix  [0]{URL }%
\providecommand \Eprint [0]{\href }%
\providecommand \doibase [0]{https://doi.org/}%
\providecommand \selectlanguage [0]{\@gobble}%
\providecommand \bibinfo  [0]{\@secondoftwo}%
\providecommand \bibfield  [0]{\@secondoftwo}%
\providecommand \translation [1]{[#1]}%
\providecommand \BibitemOpen [0]{}%
\providecommand \bibitemStop [0]{}%
\providecommand \bibitemNoStop [0]{.\EOS\space}%
\providecommand \EOS [0]{\spacefactor3000\relax}%
\providecommand \BibitemShut  [1]{\csname bibitem#1\endcsname}%
\let\auto@bib@innerbib\@empty
\bibitem [{\citenamefont {Hamilton}(1834)}]{hamilton1834general}%
  \BibitemOpen
  \bibfield  {author} {\bibinfo {author} {\bibfnamefont {W.~R.}\ \bibnamefont
  {Hamilton}},\ }\href@noop {} {\emph {\bibinfo {title} {On a general method in
  dynamics}}}\ (\bibinfo  {publisher} {Richard Taylor},\ \bibinfo {year}
  {1834})\BibitemShut {NoStop}%
\bibitem [{\citenamefont {Heaviside}(1903)}]{heaviside1903}%
  \BibitemOpen
  \bibfield  {author} {\bibinfo {author} {\bibfnamefont {O.}~\bibnamefont
  {Heaviside}},\ }\bibfield  {title} {\bibinfo {title} {The principle of least
  action. {L}agrange's equations},\ }\href {https://doi.org/10.1038/067297b0}
  {\bibfield  {journal} {\bibinfo  {journal} {Nature}\ }\textbf {\bibinfo
  {volume} {67}},\ \bibinfo {pages} {297} (\bibinfo {year} {1903})}\BibitemShut
  {NoStop}%
\bibitem [{\citenamefont {Feynman}(1942)}]{feynman1942principle}%
  \BibitemOpen
  \bibfield  {author} {\bibinfo {author} {\bibfnamefont {R.~P.}\ \bibnamefont
  {Feynman}},\ }\bibfield  {title} {\bibinfo {title} {The principle of least
  action in quantum mechanics.},\ }\href@noop {} {\bibfield  {journal}
  {\bibinfo  {journal} {Ph. D. Thesis}\ } (\bibinfo {year} {1942})}\BibitemShut
  {NoStop}%
\bibitem [{\citenamefont {Brenier}(1989)}]{brenier1989least}%
  \BibitemOpen
  \bibfield  {author} {\bibinfo {author} {\bibfnamefont {Y.}~\bibnamefont
  {Brenier}},\ }\bibfield  {title} {\bibinfo {title} {The least action
  principle and the related concept of generalized flows for incompressible
  perfect fluids},\ }\href@noop {} {\bibfield  {journal} {\bibinfo  {journal}
  {Journal of the American Mathematical Society}\ }\textbf {\bibinfo {volume}
  {2}},\ \bibinfo {pages} {225} (\bibinfo {year} {1989})}\BibitemShut {NoStop}%
\bibitem [{\citenamefont {Rojo}\ and\ \citenamefont
  {Bloch}(2018)}]{rojo2018principle}%
  \BibitemOpen
  \bibfield  {author} {\bibinfo {author} {\bibfnamefont {A.}~\bibnamefont
  {Rojo}}\ and\ \bibinfo {author} {\bibfnamefont {A.}~\bibnamefont {Bloch}},\
  }\href@noop {} {\emph {\bibinfo {title} {The principle of least action:
  History and Physics}}}\ (\bibinfo  {publisher} {Cambridge University Press},\
  \bibinfo {year} {2018})\BibitemShut {NoStop}%
\bibitem [{\citenamefont {Carcassi}\ and\ \citenamefont
  {Aidala}(2022)}]{aop-phys-ReversePhysics}%
  \BibitemOpen
  \bibfield  {author} {\bibinfo {author} {\bibfnamefont {G.}~\bibnamefont
  {Carcassi}}\ and\ \bibinfo {author} {\bibfnamefont {C.~A.}\ \bibnamefont
  {Aidala}},\ }\bibfield  {title} {\bibinfo {title} {Reverse physics: From laws
  to physical assumptions},\ }\href
  {https://doi.org/10.1007/s10701-022-00555-z} {\bibfield  {journal} {\bibinfo
  {journal} {Foundations of Physics}\ }\textbf {\bibinfo {volume} {52}},\
  \bibinfo {pages} {1} (\bibinfo {year} {2022})}\BibitemShut {NoStop}%
\bibitem [{\citenamefont {Carcassi}\ \emph {et~al.}(2018)\citenamefont
  {Carcassi}, \citenamefont {Aidala}, \citenamefont {Baker},\ and\
  \citenamefont {Bieri}}]{aop-phys-blueprint}%
  \BibitemOpen
  \bibfield  {author} {\bibinfo {author} {\bibfnamefont {G.}~\bibnamefont
  {Carcassi}}, \bibinfo {author} {\bibfnamefont {C.~A.}\ \bibnamefont
  {Aidala}}, \bibinfo {author} {\bibfnamefont {D.~J.}\ \bibnamefont {Baker}},\
  and\ \bibinfo {author} {\bibfnamefont {L.}~\bibnamefont {Bieri}},\ }\bibfield
   {title} {\bibinfo {title} {From physical assumptions to classical and
  quantum {H}amiltonian and {L}agrangian particle mechanics},\ }\href
  {https://doi.org/10.1088/2399-6528/aaba25} {\bibfield  {journal} {\bibinfo
  {journal} {Journal of Physics Communications}\ }\textbf {\bibinfo {volume}
  {2}},\ \bibinfo {pages} {045026} (\bibinfo {year} {2018})}\BibitemShut
  {NoStop}%
\bibitem [{\citenamefont {Souriau}\ and\ \citenamefont
  {Cushman}(1970)}]{souriau1970structure}%
  \BibitemOpen
  \bibfield  {author} {\bibinfo {author} {\bibfnamefont {J.-M.}\ \bibnamefont
  {Souriau}}\ and\ \bibinfo {author} {\bibfnamefont {C.}~\bibnamefont
  {Cushman}},\ }\href@noop {} {\emph {\bibinfo {title} {Structure des systèmes
  dynamiques: maîtrises de mathématiques}}}\ (\bibinfo  {publisher} {Dunod
  Universit\'{e}},\ \bibinfo {year} {1970})\BibitemShut {NoStop}%
\bibitem [{\citenamefont {Abraham}\ and\ \citenamefont
  {Marsden}(1978)}]{abraham1978foundations}%
  \BibitemOpen
  \bibfield  {author} {\bibinfo {author} {\bibfnamefont {R.}~\bibnamefont
  {Abraham}}\ and\ \bibinfo {author} {\bibfnamefont {J.}~\bibnamefont
  {Marsden}},\ }\href@noop {} {\emph {\bibinfo {title} {Foundations of
  Mechanics 2nd edn}}}\ (\bibinfo  {publisher} {Westview Press},\ \bibinfo
  {year} {1978})\BibitemShut {NoStop}%
\bibitem [{\citenamefont {Arnold}(1989)}]{arnold1989mathematical}%
  \BibitemOpen
  \bibfield  {author} {\bibinfo {author} {\bibfnamefont {V.~I.}\ \bibnamefont
  {Arnold}},\ }\href@noop {} {\emph {\bibinfo {title} {Mathematical Methods of
  Classical Mechanics}}},\ \bibinfo {edition} {2nd}\ ed.\ (\bibinfo
  {publisher} {Springer},\ \bibinfo {year} {1989})\BibitemShut {NoStop}%
\bibitem [{\citenamefont {Marsden}\ and\ \citenamefont
  {Ratiu}(1999)}]{marsden1999introduction}%
  \BibitemOpen
  \bibfield  {author} {\bibinfo {author} {\bibfnamefont {J.~E.}\ \bibnamefont
  {Marsden}}\ and\ \bibinfo {author} {\bibfnamefont {T.~S.}\ \bibnamefont
  {Ratiu}},\ }\href@noop {} {\emph {\bibinfo {title} {Introduction to Mechanics
  and Symmetry}}}\ (\bibinfo  {publisher} {Springer},\ \bibinfo {year}
  {1999})\BibitemShut {NoStop}%
\bibitem [{\citenamefont {Lanczos}(1949)}]{lanczos1949variational}%
  \BibitemOpen
  \bibfield  {author} {\bibinfo {author} {\bibfnamefont {C.}~\bibnamefont
  {Lanczos}},\ }\href@noop {} {\emph {\bibinfo {title} {The variational
  principles of mechanics}}}\ (\bibinfo  {publisher} {University of Toronto
  Press},\ \bibinfo {year} {1949})\BibitemShut {NoStop}%
\bibitem [{\citenamefont {Synge}(1960)}]{synge1960encyclopedia}%
  \BibitemOpen
  \bibfield  {author} {\bibinfo {author} {\bibfnamefont {J.}~\bibnamefont
  {Synge}},\ }\href@noop {} {\bibinfo {title} {Encyclopedia of physics vol 3/1
  ed. {S}. {F}l{\"u}gge}} (\bibinfo {year} {1960})\BibitemShut {NoStop}%
\bibitem [{\citenamefont {Peliti}(2011)}]{peliti2011statistical}%
  \BibitemOpen
  \bibfield  {author} {\bibinfo {author} {\bibfnamefont {L.}~\bibnamefont
  {Peliti}},\ }\href@noop {} {\emph {\bibinfo {title} {Statistical mechanics in
  a nutshell}}}\ (\bibinfo  {publisher} {Princeton University Press},\ \bibinfo
  {year} {2011})\BibitemShut {NoStop}%
\bibitem [{\citenamefont {Carcassi}\ and\ \citenamefont
  {Aidala}(2021)}]{aop-book}%
  \BibitemOpen
  \bibfield  {author} {\bibinfo {author} {\bibfnamefont {G.}~\bibnamefont
  {Carcassi}}\ and\ \bibinfo {author} {\bibfnamefont {C.~A.}\ \bibnamefont
  {Aidala}},\ }\href {https://doi.org/10.3998/mpub.12204707} {\emph {\bibinfo
  {title} {Assumptions of Physics}}}\ (\bibinfo  {publisher} {Michigan
  Publishing},\ \bibinfo {year} {2021})\BibitemShut {NoStop}%
\end{thebibliography}%

\section*{Supplementary Information}

In this Supplementary Information we will review the arguments in more detail. Given that this article is still aimed at a physics audience, the focus will be conceptual rigor, rather than mathematical rigor. What is important is to establish a precise map between physical concepts and their mathematical representation, so that it is clear what physical requirement is captured by each mathematical definition.

We divide the material into three sections: the first will address the single d.o.f. case, the second will give a geometric interpretation of the tools of differential topology that is suitable for physicists, and lastly we address the multiple d.o.f. case.

\subsection*{One degree of freedom}

For the case of one degree of freedom we are going to use vector calculus. This is possible since the problem can be studied with three variables (position, momentum and time)
\begin{equation}
\tag*{(\ref{sdof_variables})}
	\xi^a = [ q, p, t]
\end{equation}
which is the only setting where vector calculus truly works. In this sense, nature gave us a hand as three dimensional spaces are much more intuitive for us, and we can therefore develop the ideas in a familiar context before generalizing.

We break down the problem into sequential steps, which serves as a summary for the whole argument. We will have to:
\begin{enumerate}[label=(\roman*)]
	\item show that assuming determinism and reversibility \ref{assum_detrev} is equivalent to assuming the existence of the displacement field $\vec{S}$ defined in \ref{sdof_displacement} and its being divergence-free as in \ref{sdof_div_free}
	\item show that the vector potential $\vec{\theta}$ for $\vec{S}$ defined in \ref{sdof_displacement_potential} can be expressed, without loss of generality, as \ref{sdof_potential_expression}
	\item show that \ref{sdof_div_free} and \ref{sdof_potential_expression} recover Hamilton's equations as in \ref{sdof_Ham_eq}
	\item show that the flow of $\vec{S}$ through a surface $\Sigma$ delimited by a path $\gamma$ and its variation $\gamma'$ is the variation of the integral of the vector potential, as claimed in \ref{sdof_action}
	\item show that the integral of the vector potential is stationary for and only for the solutions of the equations of motion
	\item show that the integral of the vector potential reduces to the action if and only if \ref{assum_kineq} applies.
\end{enumerate}

For (i), determinism and reversibility means that giving the state at some point in time means constraining the state at all times, that the system can evolve in only one way. In math terms, if $X$ is the manifold of all possible states at all times, given $P = \{q_0, p_0, t_0\}$ there will be only one possible evolution $\gamma : \mathbb{R} \to X$ such that $\gamma(t_0) = P$.

If $X$ were a discrete space, the previous mathematical condition would be sufficient, but on a manifold it is not. Consider a damped harmonic oscillator: it satisfies the previous condition, yet the system will concentrate at the equilibrium. What needs to happen is that densities also need to be mapped exactly. That is, if a certain fraction of the system is in a given initial state, the same fraction of the system must be in the corresponding final state. Equivalently, if the system has a given probability of starting in a given initial state, it must have the same probability of ending in the corresponding final state. In mathematical terms, if $\rho : X \to \mathbb{R}$ is a statistical distribution (or a probability density) defined on our extended phase space and $\gamma$ is a possible evolution, we must have $\rho(\gamma(t_0)) = \rho(\gamma(t_1))$.

This requirement tells us two things. First, that $X$ must be a differentiable manifold. To define densities over a set of variables and map them backward and forward in time Jacobians must be defined, which means the derivative across state variables must be defined as well, in particular with respect to time, which plays the double role of a coordinate of $X$ and the evolution parameter. This means that the evolutions must be differentiable curves and therefore the displacement $\vec{S}$ must be defined, which leads to
\begin{equation}
\tag*{(\ref{sdof_displacement})}
	\begin{aligned}
		\vec{S} &= \left[ \frac{dq}{dt},\frac{dp}{dt},\frac{dt}{dt} \right] \\
		&= S^a e_a = \frac{d\xi^a}{dt} e_a .
	\end{aligned}
\end{equation}
In the literature of dynamical systems, the displacement vector field $\vec{S}$ is simply called the vector field of the dynamical system. The expression of the vector field is what determines the equations of motion in a dynamical system. The double role of time gives us
\begin{equation}
\tag*{(\ref{sdof_time_constraint})}
 	S^t=\frac{dt}{dt}=1.
 \end{equation}

Secondly, since the densities must be constant over evolutions, the flow of these densities through a closed surface in $X$ must be zero: each evolution will enter and exit the region delimited by the surface, giving no net contribution. Mathematically this requires the displacement to be divergence-free, which leads to
\begin{equation}
\tag*{(\ref{sdof_div_free})}
	\nabla \cdot \vec{S} = \partial_a S^a = 0.
\end{equation}

For (ii), since $\vec{S}$ is divergence-free, we can find a vector potential
\begin{equation}
\tag*{(\ref{sdof_displacement_potential})}
	\begin{aligned}
		\vec{\theta} &= [\theta_q, \theta_p, \theta_t] \\
		&= \theta_a e^a \\
		\vec{S} &= - \nabla \times \vec{\theta} \\
		&= - \epsilon^{abc} \partial_b \theta_c \, e_a. \\
	\end{aligned}
\end{equation}
Note that the vector potential is not uniquely identified, since $\nabla \times(\vec{\theta} + \nabla f) = \nabla \times \vec{\theta}$ and therefore the displacement field $\vec{S}$ remains unchanged. This is commonly referred to as gauge freedom in physics, and it is analogous to what happens for a magnetic field or for an incompressible fluid.

We use this gauge freedom to set the momentum component to zero. In fact, we can always choose $f(q,p,t)$ such that $\partial_p f = -\theta_p$ and redefine $\vec{\theta}$ to be $\vec{\theta} + \nabla f$. Therefore we can set
\begin{equation}
	\vec{\theta} = [\theta_q, 0, \theta_t]
\end{equation}
without loss of generality.

We now use constraint \ref{sdof_time_constraint} on the time component. We have
\begin{align}
	S^t = \frac{dt}{dt} &= - \left(\frac{\partial}{\partial q}  \theta_p - \frac{\partial}{\partial p}  \theta_q\right) \\
	&= - \left(\frac{\partial}{\partial q}  0 - \frac{\partial}{\partial p}  \theta_q\right) \\
	& = \frac{\partial \theta_q}{\partial p} = 1
\end{align}
Integrating, we have $\theta_q = p + g(q,t)$ where $g(q,t)$ is an arbitrary function. This function can be set to zero without loss of generality as it can be removed with a gauge transformation where $f(q,t)$ does not depend on $p$, and therefore $\theta_p$ will remain unchanged. Therefore we have:
\begin{equation}
	\vec{\theta} = [p, 0, \theta_t]
\end{equation}
Lastly, we rename the last component as $\theta_t = -H$ which leads to 
\begin{equation}
\tag*{(\ref{sdof_potential_expression})}
	\begin{aligned}
		\vec{\theta} &= [p, 0, -H] \\
		&= p e^q - H e^t.
	\end{aligned}
\end{equation}

The potential $\vec{\theta}$ is closely related to the canonical one-form of symplectic geometry and the contact form of contact geometry, but there are some differences. Both the canonical one-form and the contact form are a specific covector, defined with specific coordinates. For us, $\vec{\theta}$ is a potential and therefore is not fully specified by the physics. Writing down a specific potential means choosing a gauge, therefore our potential corresponds to the canonical one-form or the contact form only if a specific gauge is fixed.

For (iii), we expand each component of \ref{sdof_displacement_potential} using \ref{sdof_potential_expression}. Again note the double role of time: $\frac{dp}{dt}$ is not the same as $\frac{\partial p}{\partial t}$.\footnote{We could be more precise and use $\tau$ for the evolution parameter, leaving $t$ just for the coordinate. The treatment would automatically acquire a relativistic flavor and would enlighten us to other interesting results, though it would distract from the main focus of this paper, the action principle. } In the first case we are taking a total derivative along the evolution, and therefore the momentum can change. In the second case we are taking a partial derivative at constant $q$ and $p$ by definition, and therefore $\frac{\partial p}{\partial t}=0$. We have
\begin{equation}
\begin{aligned}
	S^q = \frac{dq}{dt}
	&= - \left( \frac{\partial}{\partial p} \theta_t - \frac{\partial}{\partial t} \theta_p \right) \\
	&= - \left( \frac{\partial}{\partial p} (-H) - \frac{\partial}{\partial t} 0 \right) \\
	& = \frac{\partial H}{\partial p}
\end{aligned}
\end{equation}
\begin{equation}
\begin{aligned}
	S^p = \frac{dp}{dt}
	&= - \left( \frac{\partial}{\partial t} \theta_q - \frac{\partial}{\partial q} \theta_t \right) \\
	&= - \left( \frac{\partial}{\partial t} p - \frac{\partial}{\partial q} (-H) \right) \\
	& = - \frac{\partial H}{\partial q}
\end{aligned}
\end{equation}
\begin{equation}
\begin{aligned}
	S^t = \frac{dt}{dt}
	&= - \left( \frac{\partial}{\partial q} \theta_p - \frac{\partial}{\partial p} \theta_q \right) \\
	&= - \left( \frac{\partial}{\partial q} 0 - \frac{\partial}{\partial p} p \right) \\
	& = 1
\end{aligned}
\end{equation}
We have recovered Hamilton's equations
\begin{equation}
\tag*{(\ref{sdof_Ham_eq})}
\begin{aligned}
	\frac{dq}{dt} &= \frac{\partial H}{\partial p} \\
	\frac{dp}{dt} &=-\frac{\partial H}{\partial q}
\end{aligned}
\end{equation}
as the equations for a deterministic and reversible system. The only thing that the equations say is that states move in time at the same rate (i.e. $\frac{dt}{dt} = 1$) with an incompressible flow (i.e. no states are created or destroyed). That is the whole physical and geometric content of those equations.

Note that if we started from Hamilton's equations, the system would be deterministic and reversible: position and momentum at a fixed time would indeed allow us to pick a single differentiable evolution and, by Liouville's theorem, densities would be constant along such evolution. Therefore we have an equivalence between assumption \ref{assum_detrev} and Hamilton's equations \ref{sdof_Ham_eq} in the single degree of freedom case.

Now that we have developed a clear geometric and physical understanding of Hamilton's equations, we can concentrate on the action principle. For (iv) we can apply Stokes' theorem. We take a path $\gamma$ and a variation $\gamma'$. Since they share the endpoints, taken together they form a closed curve where $\gamma'$ is taken with the opposite orientation. Because $\vec{S}$ is divergence-free, the flow of $\vec{S}$ through a surface depends only on its contour. Moreover, by Stokes' theorem, the flow through the surface will be equal to the integral of the potential over the contour. Thus we have
\begin{equation}
\tag*{(\ref{sdof_action})}
\begin{aligned}
	- \iint_{\Sigma} \vec{S} \cdot d\vec{\Sigma} &=
	\iint_{\Sigma} \left( \nabla \times \vec{\theta} \right) \cdot d\vec{\Sigma} \\
	&= \oint_{\partial \Sigma} \vec{\theta}  \cdot d\vec{\gamma} \\
	&= \int_{\gamma} \vec{\theta} \cdot d\vec{\gamma} - \int_{\gamma'} \vec{\theta} \cdot d\vec{\gamma}' \\
	&= \delta \int_{\gamma} \vec{\theta} \cdot d\vec{\gamma}.
\end{aligned}
\end{equation}

For (v), note that in the neighborhood of $\gamma$ the infinitesimal surface element $d\vec{\Sigma}$ has one side along $\gamma$ itself. Therefore we can write $d\vec{\Sigma} = d\vec{\gamma} \times d\vec{\lambda}$, where $d\vec{\lambda}$ is the infinitesimal displacement that connects $\gamma$ with $\gamma'$. The integrand then becomes the triple product
\begin{equation}
	\vec{S} \cdot d\vec{\gamma} \times d\vec{\lambda}.
\end{equation}
We can understand it geometrically as the volume element given by the displacement $\vec{S}$, the infinitesimal line element $d\vec{\gamma}$ and the infinitesimal variation $d\vec{\lambda}$ that moves $\gamma$ to $\gamma'$.

If $\gamma$ is a solution of the equations of motion, $\gamma$ is a field line of $\vec{S}$: $\gamma$ is tangent to $\vec{S}$ at all points. We can write $d\vec{\gamma} = \vec{S} dt$. The triple product, in this case, is zero regardless of what $d\vec{\lambda}$ is. We can reason the other way: we look for a line $\gamma$ such that the triple product is zero regardless of what $d\vec{\lambda}$ is. The only way for that to happen is if $\gamma$ is tangent to $\vec{S}$ at each point, which means $\gamma$ is a solution of the equations of motion. This means that the paths that make the line integral of the vector potential stationary are those and only those that solve the equations of motion. This recovers the principle of stationary action in Hamiltonian form
\begin{equation}
\tag*{(\ref{sdof_stationary_action})}
	0 =\delta \int_{\gamma} \vec{\theta} \cdot d\vec{\gamma} = - \iint_{\Sigma} \vec{S} \cdot d\vec{\Sigma} 
\end{equation}

We want to stress that this variational principle applies to any divergence-free field. For example, for a magnetic field we could write:
\begin{equation}
	\iint_{\Sigma} \vec{B} \cdot d\vec{\Sigma} = \delta \int_{\gamma} \vec{A} \cdot d\vec{\gamma}.
\end{equation}
We would find that the field lines of $\vec{B}$ are the lines and only the lines for which the integral of the vector potential is stationary. Similarly, we would find that the streamlines of an incompressible fluid are the lines and only the lines for which the integral of the vector potential of the flow is stationary. While these properties may be of limited practical use, the geometric insight is the same.

For (vi), if the assumption \ref{assum_kineq} of kinematic equivalence  holds, we can from a trajectory in space $q(t)$ reconstruct the state at any time. That is, if we are given $q(t)$ we can reconstruct the full evolution $\gamma(t)$ in the extended phase space.

We saw before that the evolution must be differentiable, which means $q(t)$ must also be differentiable. Therefore the velocity $\dot{q} = \frac{dq}{dt}$ is defined at every instant. Using assumption \ref{assum_detrev}, we know that each evolution is fully identified by position and momentum at a given time. Therefore the space of all possible evolutions is two-dimensional. Given that for each evolution $\gamma(t)$ there is one and only one spatial trajectory $q(t)$, the space of all spatial trajectories must also be two-dimensional and therefore fully identified by position and velocity at a given instant. Therefore there must be a map $p(q, \dot{q}, t)$ that allows us to reconstruct the momentum at a given time by knowing position and velocity at that time.

We can use this map to express the Hamiltonian form of the principle of stationary action only in terms of position and velocity. We have
\begin{equation}
\tag*{(\ref{sdof_Lagrangian})}
	\begin{aligned}
		\delta \int_\gamma \vec{\theta} \cdot d\vec{\gamma} &= \delta \int^{t_2}_{t_1} \vec{\theta} \cdot \frac{d\vec{\gamma}}{dt} dt \\
		&= \delta \int^{t_2}_{t_1} \left(\theta_q \frac{dq}{dt} + \theta_p \frac{dp}{dt} + \theta_t \frac{dt}{dt}\right) dt \\
		&= \delta\int^{t_2}_{t_1} \left(p \frac{dq}{dt} + 0 \frac{dp}{dt} - H \frac{dt}{dt}\right) dt \\
		&= \delta \int^{t_2}_{t_1} \left(p \frac{dq}{dt} - H\right) dt \\
		&= \delta \int^{t_2}_{t_1} L(q, \dot{q}, t) \, dt = 0
	\end{aligned}
\end{equation}
We recognize the principle of stationary action in standard form.

Note that the argument can proceed in the opposite way. If we have a system for which the principle of stationary action yields a unique solution, then we can write $p=\frac{\partial L}{\partial \dot{q}} = p(q, \dot{q}, t)$ which means we are able to reconstruct the momentum, and therefore the state, at each time. The system satisfies \ref{assum_kineq}. Moreover, we can find a unique Hamiltonian such that the solutions of the equations of motion are exactly those paths that satisfy the principle of stationary action. Therefore the system satisfies \ref{assum_detrev} as well.

Before proceeding to the generalization, a few observations are in order.
\begin{itemize}
	\item We stress that the vector potential $\vec{\theta}$ is not uniquely defined, and therefore neither are the action nor the Lagrangian. Giving precise physical characterizations of those two objects, then, is in our view a hopeless task.
	\item We found that the Hamiltonian $H$ is the time component of the vector potential $\vec{\theta}$, not a scalar. This relativistic aspect emerges naturally out of the assumptions.
	\item One can, in principle, write Lagrangian systems for which the action principle does not yield a single solution (e.g. for $L=0$ all paths are stationary). Consistently, assumption \ref{assum_detrev} fails.
\end{itemize}

\subsection*{Introduction to differential topology}

The generalization to multiple d.o.f will require the use of differential topology. These tools are not typically taught in a standard physics curriculum. If they are, their presentation is dense and abstract.
Apart from the abstract definitions, the notation gets in the way of the physics. For example, writing $\frac{\partial}{\partial x^i}$ for the vector basis makes sense if one is studying the space of linear operators defined by derivations. When studying the space of velocities, however, writing $\vec{v} = v^i\frac{\partial}{\partial x^i}$ decreases understanding. Therefore we will use a notation and nomenclature that is closer to what physicists already use (e.g. Einstein notation) and need to picture. For example, we will use $e_i$ for vector basis and $e^i$ for covector (dual) basis. To those that believe we should just use the ``correct'' mathematical nomenclature, we retort that there is no such thing. Notations and names are chosen within a certain context to make the points of the discourse more clear.

What we will give here, then, is not a fully rigorous introduction to the subject, but rather a geometric understanding of what these tools describe and a physical motivation as to why they are useful. We hope this will make it easier to follow the next section and to build a geometric and physical intuition of the principle of stationary action, which is the focus of this article.



The core objective of using differential topology in physics is to characterize those quantities that are associated with a region of space and can be understood as sums of independent contributions. The mass $m$ within a region $V$ can be understood as the sum of the contributions of a mass density $\rho$ within each infinitesimal volume $dV$:
\begin{equation*}
	m(V) = \iiint_V \rho dV.
\end{equation*}
The magnetic flux $\Phi$ through a surface $\Sigma$ can be understood as the sum of the contributions of the magnetic field $\vec{B}$ over each infinitesimal $\vec{d\Sigma}$:
\begin{equation*}
	\Phi(\Sigma) = \iint_\Sigma \vec{B} \cdot \vec{d\Sigma}.
\end{equation*}
The work $W$ over a path $\gamma$ can be understood as the sum of the contributions of a force field $\vec{f}$ over each infinitesimal line $\vec{d\gamma}$:
\begin{equation*}
	W(\gamma) = \int_\gamma \vec{f} \cdot \vec{d\gamma}.
\end{equation*}

Though the dimensionality changes, a pattern emerges: the functionals $W$, $\Phi$ and $m$ all take a region of space while $\vec{f}$, $\vec{B}$ and $\rho$ act on infinitesimal regions. However, the vector representation is in terms of different objects (i.e. vector, pseudo-vector and pseudo-scalar) and operations. Thus it fails to provide a nice generalization.

Suppose we wrote
\begin{equation*}
\begin{aligned}
	W(\gamma) &= \int_\gamma f(d\gamma) \\
	\Phi(\Sigma) &= \iint_\Sigma B(d\Sigma) \\
	m(V) &= \iiint_V \rho(dV).
\end{aligned}
\end{equation*}
Here we don't have vectors, but functions of infinitesimal regions, which we call differential forms. The force, in this notation, is a one-form (or covector) as it takes one dimensional infinitesimal regions (i.e. vectors); the magnetic field is a two-form; the density is a three-form. We can also say that a scalar field, like the temperature, is a zero-form.

An infinitesimal region can also be understood as a parallelepiped, which is fully identified by its sides. Therefore a differential form can be understood as acting on a set of vectors, the sides of the parallelepiped, that matches the dimensionality of the form. A one-form will take one vector, a two-form two vectors and so on. All forms must be anti-symmetric because switching the order of the sides does not change the parallelepiped, just its orientation.

This picture is much easier to generalize to any dimensionality. If we note $S^k$ the space of all the $k$ dimensional subregions, a $k$-functional $F : S^k \to \mathbb{R}$ is a linear functional that takes a $k$-surface $\sigma$ and returns a number. This can be expressed as
\begin{equation*}
	F(\sigma) = \int_\sigma \omega(d\sigma),
\end{equation*}
the integral of a $k$-form $\omega : V^k \to \mathbb{R}$ that takes an infinitesimal $k$-surface $d\sigma$ (or equivalently a set of $k$ vectors) and returns a number.

This allows us to identify properties and theorems that apply to all functionals and forms regardless of their dimensionality. For example, if $\sigma_1$ and $\sigma_2$ are two disjoint regions, $F(\sigma_1 \cup \sigma_2) = F(\sigma_1) + F(\sigma_2)$, which also tells us $F(\emptyset) = 0$ for any functional. Additionally, $k$-forms are linear maps and can be written in terms of the tensor product of the co-basis $e^a$. We have:
\begin{equation*}
	\begin{aligned}
		f &= f_a e^a \\
		B &= B_{ab} e^a \otimes e^b \\
		\rho &= \rho_{abc} e^a \otimes e^b \otimes e^c.
	\end{aligned}
\end{equation*}
These tensors are all anti-symmetric as the forms are anti-symmetric.





We can see how this works for a force applied along a path. The infinitesimal displacement along $\gamma$ is given by
\begin{equation*}
	d\gamma = dx^a e_a.
\end{equation*}
Therefore 
\begin{equation*}
	\begin{aligned}
		W(\gamma) &= \int_\gamma f(d\gamma) = \int_\gamma  f_a e^a(dx^b e_b) \\
		&= \int_\gamma  f_a dx^b e^a( e_b)
		= \int_\gamma  f_a dx^b \delta^a_b
		= \int_\gamma  f_a dx^a.
	\end{aligned}
\end{equation*}

Of all generalized properties, the most impressive one is certainly the generalized Stokes' theorem. Compare the divergence (or Gauss') theorem:
\begin{equation*}
	\iiint_V (\nabla \cdot \vec{B}) \, dV = \oiint_{\Sigma = \partial V} \vec{B} \cdot \vec{d\Sigma}
\end{equation*}
with Stokes' theorem
\begin{equation*}
	\iint_\Sigma (\nabla \times \vec{f}) \cdot \vec{d\Sigma} = \oint_{\gamma = \partial \Sigma} \vec{f} \cdot \vec{d\gamma}
\end{equation*}
and the divergence theorem
\begin{equation*}
	\int_\gamma (\nabla \varphi) \cdot \vec{d\gamma} = \left[ \varphi \right]_A^B .
\end{equation*}
In all three cases we have a region where we define the integral of some differential operator applied to a field. This is equated to the integral over the boundary of the field itself. But again, the vector calculus notation prevents us from generalizing as the differential operator is somewhat different in each of the three cases.

To understand conceptually that a generalization must exist, suppose we have a $k$-functional $F$ and its corresponding $k$-form $\omega$. By definition $F$ acts on $k$-surfaces. Now suppose we have a $k+1$-dimensional surface $\Sigma$. This is not a valid argument for $F$. However, its boundary $\partial \Sigma$ will be a $k$-dimensional surface, which is a valid argument for $F$. Therefore from any $F$ we can create a $k+1$-functional, which we can call the exterior functional of $F$, noted $\partial F$, such that $\partial F(\Sigma) = F(\partial \Sigma)$. If we look at all three cases above, that is exactly what is happening: the gradient, the curl and the divergence lead to a functional one dimension higher by applying the original operation on the boundary.

Now, since every functional has a corresponding form, $\partial F$ must have a corresponding $k+1$-form which must be fully identified by $\omega$. We note it as $\partial \wedge \omega$ and write
\begin{equation*}
	\partial F(\Sigma) = \int_\Sigma \partial \wedge \omega(d\Sigma) = \oint_{\sigma = \partial \Sigma} \omega(d\sigma) = F(\partial \Sigma).
\end{equation*}
This is the generalized Stokes' theorem and $\partial \wedge \omega$ is the exterior derivative of $\omega$. We use $\partial \wedge$ as we leave $d$ reserved for the infinitesimal regions.

With a few steps, one can see that the divergence, curl and gradient correspond to the exterior derivative in three dimensions for a zero-form, one-form and two-form respectively, and that the three theorems of vector calculus are particular instances of this generalized one.

Moving on, the identities
\begin{equation*}
	\begin{aligned}
		\nabla \cdot \nabla \times \vec{f} &= 0 \\
		\nabla \times \nabla \varphi &= 0
	\end{aligned}
\end{equation*}
both correspond to the single identity
\begin{equation*}
	\partial \wedge \partial \wedge \omega = 0,
\end{equation*}
which states that the exterior derivative applied twice is also zero. This identity can be understood in terms of the functionals. We have  $\partial \partial F(\Sigma) = F(\partial \partial \Sigma)$ and the boundary of a boundary $\partial \partial \Sigma = \emptyset$ is always the empty set. So, no matter what $F$ or $\Sigma$ is, we are evaluating a functional on an empty set, which is zero.

In the same way that a divergence-free field can be written as the curl of a vector potential, or a curl-free field can be written as the gradient of a scalar potential, a form whose exterior derivative is zero $\partial \wedge \omega = 0$ (a closed form), can be written as the exterior derivative $\omega = \partial \wedge \theta$ of a lower dimensional form, at least in some neighborhood. In terms of functionals, this corresponds to the case where $F(\sigma)=0$ for all contractible surfaces, those surfaces that can be morphed to a point with a continuous transformation. If $F(\sigma)=0$ over all surfaces, instead, then the corresponding form is exact. 

The overall point here is that understanding differential topology in terms of functionals of regions gives us much better motivation and intuition for the math.

\subsection*{Multiple degrees of freedom}

The case of multiple degrees of freedom proceeds in the same way as the single degree of freedom. The main difference is that we will work with a system described by $N$ degrees of freedom, which is therefore described by the $2N+1$ variables
\begin{equation}
\tag*{(\ref{mdof_variables})}
	\xi^a = [ q^i, p_i, t].
\end{equation}
We will have to:
\begin{enumerate}[label=(\roman*)]
	\item show that assuming determinism and reversibility \ref{assum_detrev} and independence of degrees of freedom \ref{assum_indep} is equivalent to assuming the existence of the displacement field $\vec{S}$ defined by \ref{mdof_displacement} and the closed counting form $\omega$ defined by \ref{mdof_form} such that $\omega$ is killed only by the displacement field $\vec{S}$ as in \ref{mdof_displacement_kills} or by a field everywhere tangent to $\vec{S}$.
	\item show that the potential $\theta$ for $\omega$ defined in \ref{mdof_form_potential} can be expressed, without loss of generality, as \ref{mdof_potential_expression} and that $\omega$ can be expressed as $\ref{mdof_form_components}$
	\item show that \ref{mdof_form_potential}, \ref{mdof_displacement_kills} and \ref{mdof_potential_expression} recover Hamilton's equations as in \ref{mdof_Ham_eq}
	\item show that the integral of $\omega$ through a surface $\Sigma$ delimited by a path $\gamma$ and its variation $\gamma'$ is the variation of the integral of the vector potential, as claimed in \ref{mdof_action}
	\item show that the integral of the vector potential is stationary for and only for the solutions of the equations of motion
	\item show that the integral of the vector potential reduces to the action if and only if \ref{assum_kineq} applies.
\end{enumerate}

For (i), the arguments presented for the single d.o.f. still apply: \ref{assum_detrev} leads to the existence of a displacement field $\vec{S}$
\begin{equation}
\tag*{(\ref{mdof_displacement})}
	\begin{aligned}
		\vec{S} &= S^a e_a = \frac{d\xi^a}{dt} e_a =\frac{dq^i}{dt} e_{q^i} + \frac{dp_i}{dt} e_{p_i} + \frac{dt}{dt} e_t
	\end{aligned}
\end{equation}
that tells us how states move in the extended phase space. We also need to reformulate state/density conservation in time, which we cannot do simply with a surface integral of $\vec{S}$. To this end, we need a functional $\mathcal{F}$ that takes a hyper-surface $\Sigma$ in the extended phase space and returns the count of states. We have 
\begin{equation}
	\mathcal{F}(\Sigma) = \int_\Sigma \Omega(d\Sigma)
\end{equation}
where $\Omega$ is the related form. Given the one-to-one correspondence between states and evolutions under \ref{assum_detrev}, $\Omega$ can be understood as quantifying either the flow of states through the surface or the states that are present on the surface.

As in the previous case, assumption \ref{assum_detrev} tells us that the integral over a closed hyper-surface is zero, since as many states flow in as flow out. We have
\begin{equation}
	\begin{aligned}
		\oint_{\Sigma = \partial V} \Omega(d\Sigma) &= 0 = \int_V \partial \wedge \Omega(dV) \\
		\partial \wedge\Omega &= 0.
	\end{aligned}
\end{equation}
which tells us that $\Omega$ is closed.

Given that $\Omega$ quantifies the flow of $\vec{S}$ through $\Sigma$, the flow for any surface that is tangent to $\vec{S}$ must vanish. If $d\gamma = \vec{S} \varphi dt$ for some scalar $\varphi$, we must have
\begin{equation}
	\Omega(d\gamma, \cdot) = \varphi dt \Omega(\vec{S}, \cdot) = 0.
\end{equation}
The notation is a bit improper as $\vec{S}$ is not technically an infinitesimal displacement, but a rate of displacement given a parameter. However, the slight abuse of notation makes the concept more clear: the direction of motion does not contribute to the flow through the surface $\Sigma$.

On the other hand, this must be the only direction that does not contribute to the flow. Since $\vec{S}$ cannot be zero because states must flow in time, for any other direction $\vec{T}$ that is not parallel to $\vec{S}$ there must be a surface tangent to $\vec{T}$ and not $\vec{S}$. The flow through that surface cannot be zero. Therefore the only way that the flow is zero for any surface tangent to a vector $\vec{T}$ is if $\vec{T}= \varphi \vec{S}$

We now turn to assumption \ref{assum_indep}. This requires the count of states to factorize: if we have a parallelepiped in phase space, the count of states (i.e. its volume) is the product of the count of the configurations over each independent d.o.f. (i.e. the area of the sides). Equivalently, \ref{assum_indep} must allow the case of a distribution over states with no correlations between the independent d.o.f., in which case the density must factorize. Since we must be able to quantify the configuration for each d.o.f., and since each d.o.f. is two dimensional, we must have a two-form $\omega$ such that:
\begin{equation}
	\Omega = \omega^N.
\end{equation}

Using assumptions \ref{assum_detrev} and \ref{assum_indep} again we can see how $\omega$ must be closed
\begin{equation}
	\partial \wedge \omega = 0.
\end{equation}
If we imagine translating a surface along an independent variable, the number of configurations identified by the surface cannot change. This uses \ref{assum_indep}. If we imagine translating in time, again the number of configurations cannot change because of \ref{assum_detrev}. Therefore, if we imagine a parallelepiped, opposite sides must identify the same number of configurations. The integral will take the sides with an opposite orientation, and therefore there will be zero net contribution. It is interesting to note that this condition requires both assumptions.

The form $\omega$ corresponds to the symplectic form of symplectic manifolds and of the symplectic bundles of contact manifolds with the following caveats. As we are in phase space extended by time, we have a contact manifold, and $\omega$ mathematically corresponds to the exterior derivative of the contact form, which is the primitive mathematical object. Physically, this is backwards: $\omega$ is the actual primitive physical object and the contact form $\theta$ is a potential, which is unphysical as it is fixed by an arbitrary gauge. Note how the mathematical definitions make it hard to have a coherent physical interpretation. The primitive objects for contact manifolds and symplectic manifolds are different, $\theta$ in the first case and $\omega$ in the other, and they are defined and studied as two separate mathematical spaces, while physically they are the same space with or without time.

To summarize, assuming \ref{assum_detrev} and \ref{assum_indep} gives us the existence of the displacement vector field $\vec{S}$ and the closed counting two-form $\omega$ such that $\vec{S}$ identifies the only vector field that kills $\omega$.

The converse is also true: if we start with the mathematical requirements, we can recover the physical assumptions. If we are given a two-form $\omega$ on a $2N+1$ dimensional space, we can define the $2N$ form $\Omega = \omega^N$. This form will be non-zero because $\vec{S}$ is the only vector field that kills the two-form $\omega$. It will also be closed since $\partial \wedge \omega = 0$ and
\begin{equation}
	\partial \wedge \Omega = N\partial \wedge \omega \wedge \omega^{N-1} = 0.
\end{equation}
So the flow of the well defined displacement field $\vec{S}$, as measured by $\Omega$, is zero through closed surfaces. This implies \ref{assum_detrev}.

To obtain \ref{assum_indep}, if each pair of state variables $q^i$, $p_i$ forms an independent degree of freedom, only the surfaces spanned by matching position and momentum will define actual configurations, while all other combinations will not define any. Therefore:
\begin{equation}\label{canonical_conditions}
	\begin{aligned}
		\omega(e^{q^i}, e^{p_j}) = \omega_{q^i p_j} &= \delta^i_j = - \omega_{p_j q^i} \\
		\omega(e^{q^i}, e^{q^j}) = \omega_{q^i q^j} &= 0 \\
		\omega(e^{p_i}, e^{p_j}) = \omega_{p_i p_j} &= 0
	\end{aligned}
\end{equation}
If we start with a closed form $\omega$ that has a single degenerate direction, by Darboux's theorem we can find coordinates for the non-degenerate part of $\omega$ that satisfy the above conditions. This recovers \ref{assum_indep}.

Note that we have broken down exactly how each assumption contributes to the mathematical setting. If \ref{assum_indep} fails, then $\Omega$ cannot be understood as $\omega^N$, which means we have many fewer conditions on the motion: we only know that the total flow is conserved, but we cannot break it down along each d.o.f.. On the other hand, if \ref{assum_detrev} fails we may not have a single displacement field $\vec{S}$ and/or the form that counts states would not be closed. Nonholonomic constraints will, in general, make one of the assumptions fail, for example by introducing a dependence between degrees of freedom. This type of exact mapping between assumptions and their mathematical expression is one of the key goals of Reverse Physics in particular and our overall project Assumptions of Physics more in general.

For (ii), we note that since the form is closed, we can find, at least in a contractible region, a covector potential $\theta$ such that $\omega=\partial \wedge \theta$. Those familiar with symplectic and contact geometry will know that using Darboux's theorem, we can find a set of coordinates such that we can express $\theta$ as
\begin{equation}
	\tag*{(\ref{mdof_potential_expression})}
	\theta = p_i e^{q^i} - H e^t.
\end{equation}
Just stating this is, of course, not very insightful. To gain more insight, we provide the sketch of a proof that, as we have done for the single degree of freedom, uses gauge transformations to fix the form of the potential. 

As a starting point, we take conditions \ref{canonical_conditions} that state mathematically that $q^i$ and $p_i$ form independent degrees of freedom. We re-express those conditions in terms of the potential $\theta$
\begin{equation}\label{canonical_potential_conditions}
\begin{aligned}
	\omega(e^{q^i}, e^{p_j}) &= (-\partial \wedge\theta)_{q^i p_j} = -(\partial_{q^i}\theta_{p_j} - \partial_{p_j}\theta_{q^i}) = \delta^i_j \\
	\omega(e^{q^i}, e^{q^j}) &= (-\partial \wedge\theta)_{q^i q^j} = -(\partial_{q^i}\theta_{q^j} - \partial_{q^j}\theta_{q^i}) = 0 \\
	\omega(e^{p_i}, e^{p_j}) &= (-\partial \wedge\theta)_{p_i p_j} = -(\partial_{p_i}\theta_{p_j} - \partial_{p_j}\theta_{p_i}) = 0
\end{aligned}
\end{equation}
We can use our gauge freedom to set $\theta_{p_1} = 0$, much in the same way we did for the simpler case. We now have $\partial_{q^1} \theta_{p_1} = 0$ and, by the first condition, $\partial_{p_1} \theta_{q^1} = 1$. Integrating, we have $\theta_{q^1} = p_1 + g(q^i, p_2, p_3, ..., t)$ where $g$ is an arbitrary function which we can set to zero. Like in the single degree of freedom, we can do this because $g$ does not depend on $p_1$, so it can be eliminated with a gauge transformation that does not change $\theta_{p_1}$. Therefore we have:
\begin{equation}
	\theta = p_1 e^{q^1} + 0 e^{p_1} + \theta_{q^2} e^{q^2} + \theta_{p_2} e^{p_2} + ... + \theta_{t} e^{t}.
\end{equation}

Note that the components for the first degree of freedom do not depend on the other degrees of freedom. That is, for all $i>1$, $\partial_{q^i} \theta_{q^1} = \partial_{p_i} \theta_{q^1} = \partial_{q^i} \theta_{p_1} = \partial_{p_i} \theta_{p_1} = 0$. But by using conditions \ref{canonical_potential_conditions}, we find that the converse is true as well: the components of all other degrees of freedom do not depend on the first. That is, for all $i>1$, $\partial_{q^1} \theta_{q^i} = \partial_{p_1} \theta_{q^i} = \partial_{q^1} \theta_{p_i} = \partial_{p_1} \theta_{p_i} = 0$.

We can then use, again, our gauge freedom with a function that does not depend on the first two variables to set $\theta_{p_2} = 0$. And, with the same reasoning, we will be able to set $\theta_{q^2} = p_2$. And then, again, find that the first two degrees of freedom do not depend on the others, etc. At the end, we will find \ref{mdof_potential_expression}.

The insight is that the independence of the degrees of freedom leads to the independent expression of the components of the covector potential. We stress again that this is a choice, and that the specific expression of the components is not physically constrained.

The components of $\partial \wedge\theta$ are
\begin{equation}
	\begin{aligned}
		(\partial \wedge\theta)_{q^i p_j} &= \partial_{q^i}\theta_{p_j} - \partial_{p_j}\theta_{q^i} = \partial_{q^i} 0 - \partial_{p_j} p_i= - \delta^i_j \\
		(\partial \wedge\theta)_{q^i q^j} &= \partial_{q^i}\theta_{q^j} - \partial_{q^j}\theta_{q^i} = \partial_{q^i}p_j - \partial_{q^j}p_i = 0 \\
		(\partial \wedge\theta)_{p_i p_j} &= \partial_{p_i}\theta_{p_j} - \partial_{p_j}\theta_{p_i} = \partial_{p_i} 0 - \partial_{p_j} 0 = 0 \\
		(\partial \wedge\theta)_{q^i t} &= \partial_{q^i}\theta_{t} - \partial_{t}\theta_{q^i} = \partial_{q^i} (-H) - \partial_{t} p_i= - \partial_{q^i} H \\
		(\partial \wedge\theta)_{p_i t} &= \partial_{p_i}\theta_{t} - \partial_{t}\theta_{p_i} = \partial_{p_i} (-H) - \partial_{t} 0= - \partial_{p_i} H \\
	\end{aligned}
\end{equation}
Therefore we recover
\begin{equation}
	\tag*{(\ref{mdof_form_components})}
	\omega_{ab}=(-\partial \wedge\theta)_{ab} = \begin{bmatrix}
		0 & \delta^j_i & \partial_{q^i} H \\
		-\delta^i_j & 0 & \partial_{p_i} H \\
		- \partial_{q^j} H & - \partial_{p_j} H & 0
	\end{bmatrix}.
\end{equation}

For (iii) we calculate $\omega(\vec{S}, \cdot ) $. We have
\begin{equation}
\begin{aligned}
	\omega(\vec{S}, \cdot )  &= S^a \omega_{ab} e^b = 0 \\
	&= (S^{q^i}\omega_{q^ib} + S^{p_i}\omega_{p_ib} + S^{t}\omega_{tb}) e^b \\
	&= (S^{q^i}\omega_{q^iq^j} + S^{p_i}\omega_{p_iq^j} + S^{t}\omega_{tq^j}) e^{q^j} + \\
	& (S^{q^i}\omega_{q^ip_j} +  S^{p_i}\omega_{p_ip_j} + S^{t}\omega_{tp_j}) e^{p_j} + \\
	& (S^{q^i}\omega_{q^it} + S^{p_i}\omega_{p_it} + S^{t}\omega_{tt}) e^t \\
	&= (-S^{p_i}\delta^i_j - S^{t}\partial_{q^j} H ) e^{q^j} + \\
	& (S^{q^i}\delta^j_i -  S^{t}\partial_{p_j} H) e^{p_j} + \\
	& (S^{q^i} \partial_{q^i} H + S^{p_i} \partial_{p_i} H) e^t \\
\end{aligned}
\end{equation}

All components must be zero, therefore we have the following three equations:
\begin{equation}
\begin{aligned}
	S^{p_j} &= - S^{t} \partial_{q^j} H \\
	S^{q^j} &= S^{t}\partial_{p_j} H \\
	S^{q^i} \partial_{q^i} H + S^{p_i} \partial_{p_i} H &= S^{t}\partial_{p_i} H \partial_{q^i} H \\
	& - S^{t} \partial_{q^i} H \partial_{p_i} H = 0
\end{aligned}
\end{equation}
Note that the last expression is not a new equation: it is identical to zero given the previous two equations. Since $S^t = 1$, we have
\begin{equation}
	\tag*{(\ref{mdof_Ham_eq})}
\begin{aligned}
	S^{q^j} &= \partial_{p_j} H \\
	S^{p_j} &= - \partial_{q^j} H 
\end{aligned}
\end{equation}
which recovers Hamilton's equations for multiple degrees of freedom.

The argument can be run in the other direction. If we have a system that obeys \ref{mdof_Ham_eq}, we can define the two-form $\omega$ using \ref{mdof_form_components}. We can verify that $\omega$ is closed, it is killed by $\vec{S}$ and recover all the starting points. Therefore we find again complete equivalence with Hamiltonian mechanics and assumptions \ref{assum_detrev} and \ref{assum_indep}.

For those familiar with general relativity, recall that the metric tensor $g_{\alpha\beta}$ defines the geometry by defining lengths and angles. Here we find something similar in the sense that the two-form $\omega_{ab}$ is a rank two tensor that defines the geometry of the space. The difference is that $\omega_{ab}$ does not define lengths and angles, but only areas of two-dimensional surfaces, which corresponds to the number of configurations on each independent degree of freedom. Mathematically, this difference is between the symmetry of $g_{\alpha\beta} = g_{\beta\alpha}$ and the anti-symmetry of $\omega_{ab} = -\omega_{ba}$. In this context, what the equations of motion say is that states move in time at the same rate (i.e. $\frac{dt}{dt} = 1$) with a deterministic and reversible flow (i.e. no states are created or destroyed on any independent degree of freedom). Nothing else.

Comparing \ref{mdof_form_components} and \ref{mdof_Ham_eq} we see that the time components $\omega_{at}$ are actually the components of $\vec{S}$. In fact, the only non-trivial components of $\omega$ are components of $\vec{S}$. These two objects, then, are not only related: they express the same information.

For a single degree of freedom, recalling that $S^t = 1$, we can describe $\omega$ with the following nine components
\begin{equation}\label{mdof_reduceto_sdof}
	\omega_{ab}=(-\partial \wedge\theta)_{ab} = \begin{bmatrix}
		0 & S^t & -S^p \\
		-S^t & 0 & S^q \\
		S^p & - S^q & 0
	\end{bmatrix}.
\end{equation}
In three dimensions, using the Levi-Civita symbol $\epsilon^{abc}$ we can write
\begin{equation}
	S^a = \epsilon^{abc} \omega_{bc}
\end{equation}
which tells us that $\vec{S}$ is the pseudo-vector corresponding to the direction perpendicular to $\omega$. This is the same relationship that we would have, for example, for the magnetic field psuedo-vector and the spatial components of the electromagnetic field in relativity. If we express $\omega$ in terms of $\vec{\theta}$, we have
\begin{equation}
	S^a = - \epsilon^{abc} \partial_b \theta_c
\end{equation}
which recovers \ref{sdof_displacement_potential}: the displacement field is minus the curl of $\omega$. Thus we can see exactly how the general case reduces to the single d.o.f. case.

For (iv), we simply need to apply Stokes' theorem. The situation is identical to the single degree of freedom, except that the flow now is not given by the surface integral of $\vec{S}$, but of the form $\omega$. The path $\gamma$ and its variation $\gamma'$ will form a closed curve where $\gamma'$ is taken with the opposite orientation. Because $\omega$ is closed, its surface integral depends only on its contour. Moreover, by Stokes' theorem, the flow through the surface will be equal to the integral of the potential over the contour. Thus we have
\begin{equation}
\tag*{(\ref{mdof_action})}
\begin{aligned}
	- \int_{\Sigma} \omega(d\Sigma) &=
	\int_{\Sigma} \partial \wedge \theta (d\Sigma) \\
	&= \oint_{\partial \Sigma = \gamma \cup \gamma'} \theta(d\gamma) \\
	&= \int_{\gamma} \theta(d\gamma) - \int_{\gamma'} \theta (d\gamma') \\
	&= \delta \int_{\gamma} \theta (d\gamma).
\end{aligned}
\end{equation}

For (v), if the path $\gamma$ is a solution to the equations of motion, it is tangent to $\vec{S}$ at every point. Therefore we can write $d\gamma= \vec{S} dt$. Each infinitesimal surface element $d\Sigma$ will be an infinitesimal parallelogram with $d\gamma$ as one side, and the other side $d\lambda$ connects the path to the variation. We have:
\begin{equation}
\tag*{(\ref{mdof_stationary_action})}
\begin{aligned}
	\delta \int_{\gamma} \theta (d\gamma) &= -\int_\Sigma \omega(d\Sigma) \\
	&= -\int_\Sigma \omega(d\gamma, d\lambda) \\
	&= -\int_\Sigma \omega(\vec{S}, d\lambda) dt = 0.
\end{aligned}
\end{equation}
The converse is also true: suppose that, for a fixed $\gamma$, the flow through $\Sigma$ is zero no matter the choice of $\gamma'$. Then at every point $\omega(d\gamma, \cdot) dt = 0$, which means $d\gamma$ kills $\omega$. But since $\vec{S}$ is the only direction that kills $\omega$ at every point, $d\gamma$ must be everywhere tangent to $\vec{S}$ and therefore $\gamma$ is a solution of the equations of motion.

For (vi), the argument is exactly the same as in the single d.o.f.. As the evolution is differentiable, the velocity $\dot{q}^i = \frac{dq^i}{dt}$ is well defined. Under \ref{assum_kineq} there exists an invertible map $p_i=p_i(q^j, \dot{q}^j, t)$ that allows us to reconstruct the momentum at a given time by knowing position and velocity. We can therefore express the principle of stationary action only in terms of position and velocity. We have
\begin{equation}
\tag*{(\ref{mdof_Lagrangian})}
	\begin{aligned}
		\delta \int_\gamma \theta (d\gamma) &= \delta \int^{t_2}_{t_1} \theta \left(\frac{d\gamma}{dt} dt\right) \\
		&= \delta \int^{t_2}_{t_1} \left(\theta_{q^i} \frac{dq^i}{dt} + \theta_{p_i} \frac{dp_i}{dt} + \theta_t \frac{dt}{dt}\right) dt \\
		&= \delta\int^{t_2}_{t_1} \left(p_i \frac{dq^i}{dt} + 0 \frac{dp_i}{dt} - H \frac{dt}{dt}\right) dt \\
		&= \delta \int^{t_2}_{t_1} \left(p_i \frac{dq^i}{dt} - H\right) dt \\
		&= \delta \int^{t_2}_{t_1} L(q^i, \dot{q}^i, t) \, dt = 0.
	\end{aligned}
\end{equation}
This recovers the principle of stationary action in standard form.

Again, the argument can proceed in the opposite direction. If we have a system for which the principle of stationary action yields a unique solution, we can write $p_i = \frac{\partial L}{\partial \dot{q}^i} = p_i(q^j, \dot{q}^j, t)$. This means that we can reconstruct the momentum from the kinematics, and therefore \ref{assum_kineq} is satisfied. Moreover, we can define a Hamiltonian $H = p_i \dot{q}^i - L$ such that the paths that satisfy the equations of motion \ref{mdof_Ham_eq} are exactly those that satisfy the principle of stationary action. Therefore the system satisfies \ref{assum_detrev} and \ref{assum_indep} as well.

The comments on the lack of strict physicality for the values of the Lagrangian and the action remain unchanged. They both depend on a choice of gauge. The comment on the centrality of the extended phase space to understand the action principle is only reinforced. Additionally, note that by changing $t$ to $ct$ expression \ref{mdof_potential_expression} becomes
\begin{equation}\label{mdof_potential_relativistic}
	\theta = p_i e^{q^i} - \frac{H}{c} e^{ct},
\end{equation}
which bears striking resemblance to the relativistic four-momentum. While it is outside the scope of this article, the idea is that, in this setting, we already have some pre-relativistic features, even though we have not introduced a metric tensor.

We hope that this result shows that it is possible to bring physical ideas and mathematical concepts much closer together in a way that is more clear and satisfying.

\end{document}